\documentclass[aps,pra,twocolumn,showpacs,notitlepage,superscriptaddress,letterpaper]{revtex4-1}
\usepackage{graphicx, amsmath, amssymb, amsfonts, pifont, bm, cancel}
\usepackage[usenames]{color}
\usepackage{subfigure}
\usepackage[normalem]{ulem} 
\usepackage{circuitikz}
\usepackage{hyperref}

\definecolor{MyDarkBlue}{rgb}{0,0,1}

\renewcommand\Re{\operatorname{Re}}
\renewcommand\Im{\operatorname{Im}}

 \newcommand{\beq}{\begin{equation}}
 \newcommand{\eeq}{\end{equation}}
 \newcommand{\bel}{\begin{align*}}
 \newcommand{\tamam}{\end{align*}}

 \newcommand{\ket}[1]{|#1\rangle}

 \newcommand{\beqa}{\begin{eqnarray}}             
 \newcommand{\eeqa}{\end{eqnarray}}               

\begin{document}

\title{Superconducting metamaterials for waveguide quantum electrodynamics}

\author{Mohammad~Mirhosseini}
\author{Eunjong~Kim}
\author{Vinicius~S.~Ferreira}
\author{Mahmoud~Kalaee}
\author{Alp~Sipahigil}
\author{Andrew~J.~Keller}
\affiliation{Kavli Nanoscience Institute and Thomas J. Watson, Sr., Laboratory of Applied Physics, California Institute of Technology, Pasadena, California 91125, USA.}
\affiliation{Institute for Quantum Information and Matter, California Institute of Technology, Pasadena, California 91125, USA.}
 
\author{Oskar~Painter}
\email{opainter@caltech.edu}
\homepage{http://copilot.caltech.edu}
\affiliation{Kavli Nanoscience Institute and Thomas J. Watson, Sr., Laboratory of Applied Physics, California Institute of Technology, Pasadena, California 91125, USA.}
\affiliation{Institute for Quantum Information and Matter, California Institute of Technology, Pasadena, California 91125, USA.}

\date{\today}

\begin{abstract}
{The embedding of tunable quantum emitters in a photonic bandgap structure enables the control of dissipative and dispersive interactions between emitters and their photonic bath. Operation in the transmission band, outside the gap, allows for studying waveguide quantum electrodynamics in the slow-light regime. Alternatively, tuning the emitter into the bandgap results in finite range emitter-emitter interactions via bound photonic states. Here we couple a transmon qubit to a superconducting metamaterial with a deep sub-wavelength lattice constant ($\lambda/60$). The metamaterial is formed by periodically loading a transmission line with compact, low loss, low disorder lumped element microwave resonators. We probe the coherent and dissipative dynamics of the system by measuring the Lamb shift and the change in the lifetime of the transmon qubit. Tuning the qubit frequency in the vicinity of a band-edge with a group index of $n_g = 450$, we observe an anomalous Lamb shift of $10$~MHz accompanied by a $24$-fold enhancement in the qubit lifetime. In addition, we demonstrate selective enhancement and inhibition of spontaneous emission of different transmon transitions, which provide simultaneous access to long-lived metastable qubit states and states strongly coupled to propagating waveguide modes. 
}
\end{abstract}
\maketitle

Cavity quantum electrodynamics (QED) studies the interaction of an atom with a single electromagnetic mode of a high-finesse cavity with a discrete spectrum \cite{Raimond:2006wd,Reiserer:2015en}. In this canonical setting, a large photon-atom coupling is achieved by repeated interaction of the atom with a single photon bouncing many times between the cavity mirrors. Recently, there has been much interest in achieving strong light-matter interaction in a cavity-free system such as a waveguide. Waveguide QED refers to a system where a chain of atoms are coupled to a common optical channel with a continuum of electromagnetic modes over a large bandwidth. Slow-light photonic crystal waveguides are of particular interest in waveguide QED because the reduced group velocity near a bandgap preferentially amplifies the desired radiation of the atoms into the  waveguide modes \cite{Yao:2009gj,Goban:2015dr,Calajo:2016fz}. Moreover, in this configuration an interesting paradigm can be achieved by placing the resonance frequency of the atom inside the bandgap of the waveguide \cite{Bykov:1975gp,Yablonovitch:1987eb, John:1990cn, Kofman:1994gi, Hood:2016ic}. In this case the atom cannot radiate into the waveguide but the evanescent field surrounding it gives rise to a photonic bound state \cite{John:1990cn}. The interaction of such localized bound states has been proposed for realizing tunable spin-exchange interaction between atoms in a chain \cite{Munro:2017ba, Douglas:2015hd}, and also for realizing effective non-local interactions between photons \cite{Shahmoon:2016kv,Douglas:2016kr}.

While achieving efficient waveguide coupling in the optical regime requires the challenging task of interfacing atoms or atomic-like systems with nanoscale dielectric structures~\cite{Vetch2010,Yu2014,Javadi2015,Bhaskar2017}, superconducting circuits provide an entirely different platform for studying the physics of light-matter interaction in the microwave regime \cite{Blais:2004kn}. Development of the field of circuit QED has enabled fabrication of fast and tunable qubits with long coherence times \cite{Koch:2007gz,Barends:2013kz,Chen:2014cwa}. Moreover, strong coupling is readily achieved in this platform due to the deep sub-wavelength transverse confinement of photons attainable in microwave waveguides and the large electrical dipole of superconducting qubits \cite{Wallraff:2004dy}.  Microwave waveguides with strong dispersion, even ``bandgaps'' in frequency, can also be simply realized by periodically modulating the geometry of a coplanar transmission line~\cite{Pozar:1998wp}. Such an approach was recently demonstrated in a pioneering experiment by Liu and Houck~\cite{Liu:2016ic}, whereby a qubit was coupled to the localized photonic state within the bandgap of a modulated coplanar waveguide (CPW).  Satisfying the Bragg condition in a periodically modulated waveguide requires a lattice constant on the order of the wavelength \cite{Joannopoulos:2011tg}, however, which translates to a device size of approximately a few centimeters for complete confinement of the evanescent fields in the frequency range suitable for microwave qubits.  Such a restriction significantly limits the scaling in this approach, both in qubit number and qubit connectivity.

An alternative approach for tailoring dispersion in the microwave domain is to take advantage of the metamaterial concept. Metamaterials are composite structures with sub-wavelength components which are designed to provide an effective electromagnetic response \cite{Smith2000,Itoh:2006uw}. Since the early microwave work, the electromagnetic metamaterial concept has been expanded and extensively studied across a broad range of classical optical sciences~\cite{Koschny2017,Alu2017,Chen2016,Genevet2017}; however, their role in quantum optics has remained relatively unexplored, at least in part due to the lossy nature of many sub-wavelength components. Improvements in design and fabrication of low-loss superconducting circuit components in circuit QED offer a new prospect for utilizing microwave metamaterials in quantum applications.  Indeed, high quality-factor superconducting components such as resonators can be readily fabricated on a chip \cite{Goppl:2008iu, Megrant:2012cd}, and such elements have been used as a tool for achieving phase-matching in near quantum-limited traveling wave amplifiers \cite{OBrien:2014fc,Macklin:2015ek,White:2015bb} and for tailoring qubit interactions in a multimode cavity QED architecture \cite{McKay:2015hn}.

\begin{figure}[t!]
\centerline{\includegraphics[width = \columnwidth]{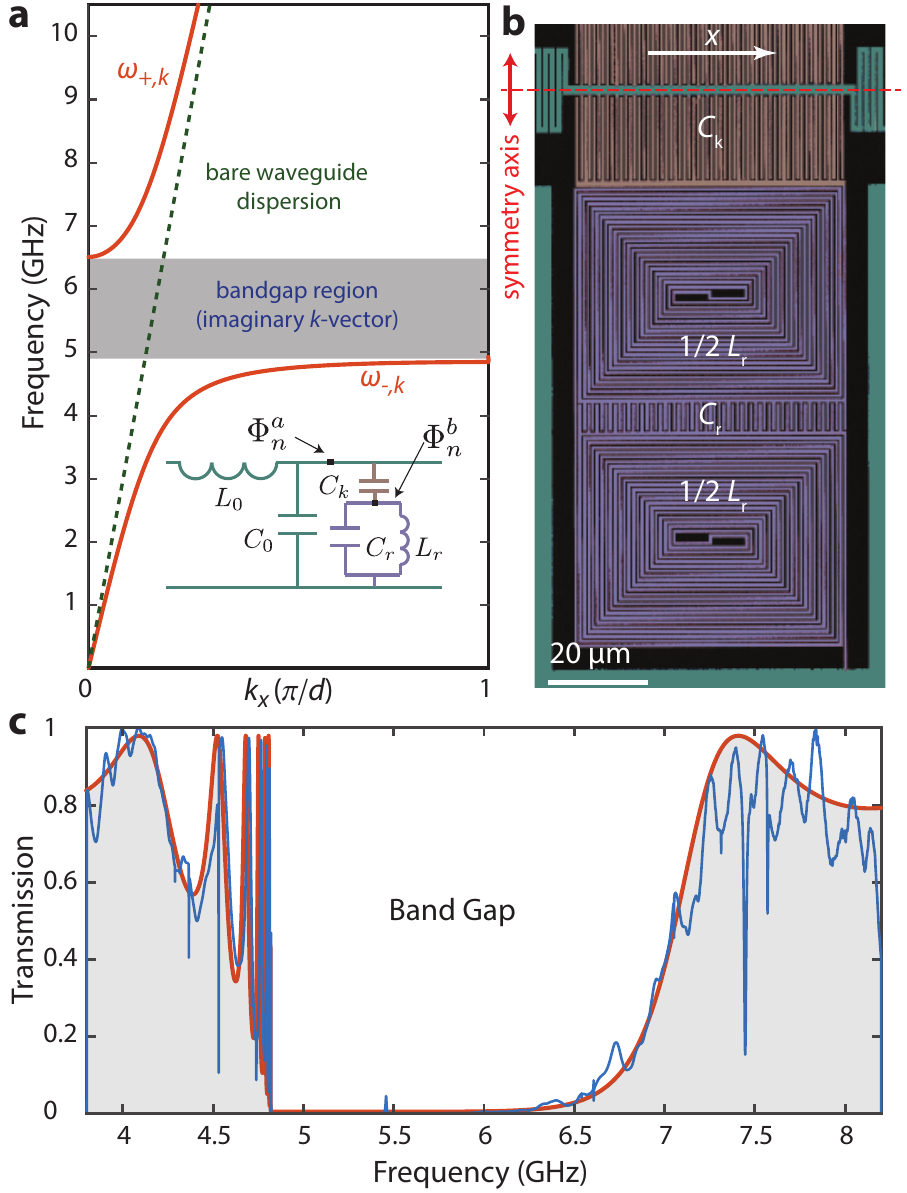}}
\caption{\textbf{Microwave metamaterial waveguide.} \textbf{a},  Dispersion relation of a CPW loaded with a periodic array of microwave resonators. The dashed line shows the dispersion relation of the waveguide without the resonators. Inset: circuit diagram for a unit cell of the periodic structure. \textbf{b}, Scanning electron microscope (SEM) image of the fabricated capacitively coupled microwave resonator with a wire width of 500 nm. The resonator region is false-colored in purple, the waveguide central conductor and the ground plane are colored green, and the coupling capacitor is shown in orange. We have used pairs of identical resonators symmetrically placed on the two sides of the transmission line to preserve the symmetry of the structure. \textbf{c}, Transmission measurement for the realized metamaterial waveguide made from 9 unit cells of resonator pairs with a wire width of 1 $\mu$m, repeated with a lattice constant of $d = 350\, \mu$m. The blue curve depicts the experimental data and the red curve shows the lumped-element model fit to the data.} 
\label{fig:Lin}
\vspace{-0.2 cm}
\end{figure}

In this paper, we utilize an array of coupled lumped-element microwave resonators to form a compact bandgap waveguide with a deep sub-wavelength lattice constant ($\lambda/60$) based on the metameterial concept. In addition to a compact footprint, these sort of structures can exhibit highly nonlinear band dispersion surrounding the bandgap, leading to exceptionally strong confinement of localized intra-gap photon states. We present the design and fabrication of such a metamaterial waveguide, and characterize the resulting waveguide dispersion and bandgap properties via interaction with a tunable superconducting transmon qubit. We measure the Lamb shift and lifetime of the qubit in the bandgap and its vicinity, demonstrating the anomalous Lamb shift of the fundamental qubit transition as well as selective inhibition and enhancement of spontaneous emission for the first two excited states of the transmon qubit.

\begin{figure*}[t]
\centerline{\includegraphics[width = \textwidth]{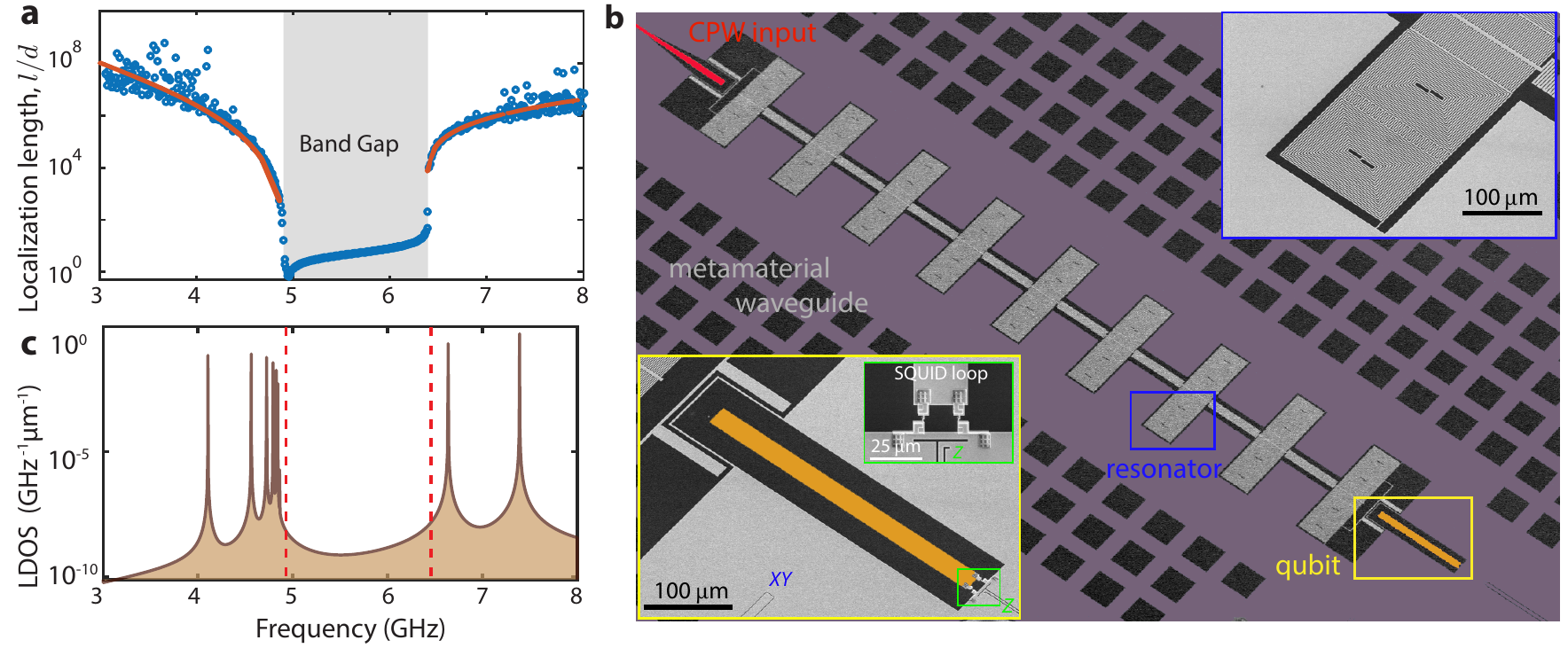}}
\caption{\textbf{Disorder effects and qubit-waveguide coupling.} \textbf{a}, Calculated localization length for a metamaterial waveguide with structural disorder and resonator loss are shown as blue dots. The waveguide parameters are determined from the fit to a lumped element model with resonator loss to the transmission data in Fig.\,\ref{fig:Lin}. Numerical simulation has been performed for $N=100$ unit cells, averaged over $10^5$ randomly realized values of the resonance frequency $\omega_{0}$, with the standard deviation $\delta\omega_0/\omega_0 = 0.5\%$. The red curve outside the gap is an analytic model based on Ref.\,\cite{HernandezHerrejon:2010ef}. \textbf{b}, SEM image of the fabricated qubit-waveguide system. The metamaterial waveguide (gray) consists of 9 periods of the resonator unit cell. The waveguide is capacitively coupled to an external CPW (red) for reflective read-out. Bottom left inset: The transmon qubit is capacitively coupled to the resonator at the end of the array. The Z drive is used to tune the qubit resonance frequency by controlling the external flux bias in the superconducting quantum interference device (SQUID) loop. The XY drive is used to coherently excite the qubit. Top right inset: capacitively coupled microwave resonator. \textbf{c}, Calculated local density of states (LDOS) at the qubit position for a metamaterial waveguide with a length of 9 unit cells and open boundary conditions. The band-edges for an infinite structure are marked with dashed red lines.} 
\label{fig:QBWGSEM}
\end{figure*}

We begin by considering the circuit model of a CPW that is periodically loaded with microwave resonators as shown in the inset to Fig.\,\ref{fig:Lin}a. The Lagrangian for this system can be constructed as a function of the node fluxes of the resonator and waveguide sections ${{\Phi}^b_n}$ and ${{\Phi}^a_n}$ \cite{Devoret:1995vn}. Assuming periodic boundary conditions and applying the rotating wave approximation, we derive the Hamiltonian for this system and find the eigenstates and energies to be (see App.~\ref{App:A}),

\begin{align}\label{Eq:ResDisp_rwa}
&\omega_{\pm,k} = \frac{1}{2} \left[ \left(\Omega_k + \omega_0 \right) \pm \sqrt{{\left( \Omega_k - \omega_0 \right)}^2 + 4  g_k^2} \right], \\
&\hat{\alpha}_{\pm,k}  = \frac{(\omega_{\pm,k} - \omega_0)}{\sqrt{ {(\omega_{\pm,k} - \omega_0)}^2 + g_k^2}}\hat{a}_{k} +  \frac{g_k}{\sqrt{ {(\omega_{\pm,k} - \omega_0)}^2 + g_k^2}}\hat{b}_{k},
\end {align}
  

\noindent where $\hat{a}_k$, and $\hat{b}_k$ describe the momentum-space annihilation operators for the bare waveguide and bare resonator sections, the index $k$ denotes the wavevector, and the parameters $\Omega_k$, ${\omega_0}$, and $g_k$ quantify the frequency of traveling modes of the bare waveguide, the resonance frequency of the microwave resonators, and coupling rate between resonator and waveguide modes, respectively.  The operators $\hat{\alpha}_{\pm,k}$ represent quasi-particle solutions of the composite waveguide, where far from the bandgap the quasi-particle is primarily composed of the bare waveguide mode, while in the vicinity of $\omega_0$ most of its energy is confined in the microwave resonators. 

Figure\,\ref{fig:Lin}a depicts the numerically calculated energy bands $\omega_{\pm,k}$ as a function of the wavevector $k$. It is evident that the dispersion has the form of an avoided crossing between the energy bands of the bare waveguide and the uncoupled resonators.  For small gap sizes, the midgap frequency is close to the resonance frequency of uncoupled resonators $\omega_0$, and unlike the case of a periodically modulated waveguide, there is no fundamental relation tying the midgap frequency to the lattice constant in this case. The form of the band structure near the higher cut-off frequency $\omega_{c+}$ can be approximated as a quadratic function $(\omega -\omega_{c+}) \propto k^2$, whereas the band structure near the lower band-edge $\omega_{c-}$ is inversely proportional to the square of the wavenumber $(\omega -\omega_{c-}) \propto 1/k^2$.  The analysis above has been presented for resonators which are capacitively coupled to a waveguide in a parallel geometry; a similar band structure can also be achieved using series inductive coupling of resonators (see App.~\ref{App:A}).

A coplanar microwave resonator is often realized by terminating a short segment of a coplanar transmission line with a length set to an integer multiple of $\lambda/4$, where $\lambda$ is the wavelength corresponding to the fundamental resonance frequency \cite{Pozar:1998wp,Goppl:2008iu,Gao:2008td}. However, it is possible to significantly reduce the footprint of a resonator by using components that mimic the behavior of lumped elements. We have used the design presented in Ref.\,\cite{Zhou:2003wf} to realize resonators in the frequency range of $6$-$10$~GHz. This design provides compact resonators by placing interdigital capacitors at the anti-nodes of the charge waves and double spiral coils near the peak of the current waves at the fundamental frequency (see Fig.\,\ref{fig:Lin}b). Further, the symmetry of the geometry results in the suppression of the second harmonic frequency and thus the elimination of an undesired bandgap at twice the fundamental resonance frequency of the band-gap waveguide.

We fabricate individual resonator pairs using an electron-beam deposited $120$~nm Al film, patterned via lift-off, on a high resistivity silicon wafer substrate of thickness $500$~$\mu$m (see Ref.~\cite{Keller2017} for further details of fabrication techniques).  In this work we have made a periodic array of 9 resonator pairs with a wire width of 1 $\mu$m and coupled them to a CPW in a periodic fashion with a lattice constant of $350$~$\mu$m to realize a metamaterial waveguide. The resonators are arranged in identical pairs placed on the opposite sides across the central waveguide conductor to preserve the symmetry of the waveguide. Figure\,\ref{fig:Lin}c shows the measured power transmission through such a finite-length metamaterial waveguide.  Here $50$-$\Omega$ CPW segments, galvanically coupled to the metamaterial waveguide, are used at the input and output ports. We find the midgap frequency of $5.83$~GHz for the structure, and a gap frequency span of $1.82$~GHz. Using the simulated value of effective refractive index of $2.54$, the midgap frequency gives a lattice constant-to-wavelength ratio of $d/\lambda \approx 1/60$.

Propagation of electromagnetic fields in the frequency range within the bandgap is exponentially attenuated with a localization length set by the imaginary part of the wavenumber. In addition, statistical variations in the electromagnetic properties of the periodic structure result in random scattering of the traveling waves in the transmission band. Such random scatterings can lead to complete trapping of propagating photons in the presence of strong disorder and an exponential extinction for weak disorder; a phenomenon known as the Anderson localization of light \cite{Wiersma:1997fp}. We have measured a random standard deviation of $0.3\%$ in the resonance frequency of the fabricated lumped-element resonators. Figure \,\ref{fig:QBWGSEM}a shows the calculated localization length as a function of frequency from numerical simulation of the independently measured disorder and loss of the resonators in the metamaterial waveguide (see App.~\ref{App:C} and ~\ref{App:D} for further details).  Near the edges of the bandgap the localization length from disorder dominates that from loss, rapidly approaching zero at the lower band-edge where the group index is largest and maintaining a large value ($6\times 10^3$ periods) at the higher band-edge where the group index is smaller. Similarly, the localization length inside the gap is inversely proportional to the curvature of the energy bands \cite{Douglas:2015hd}. Owing to the divergence (in the loss-less case) of the lower band curvature for the waveguide studied here, the localization length inside the gap approaches zero near the lower band-edge frequency as well.  These results indicate that, even with practical limitations on disorder and loss in such metamaterial waveguides, a range of photon length scales of nearly four orders of magnitude should be accessible for frequencies within a few hundred MHz of the band-edges.       

\begin{figure}[t]
\centerline{\includegraphics[width = \columnwidth]{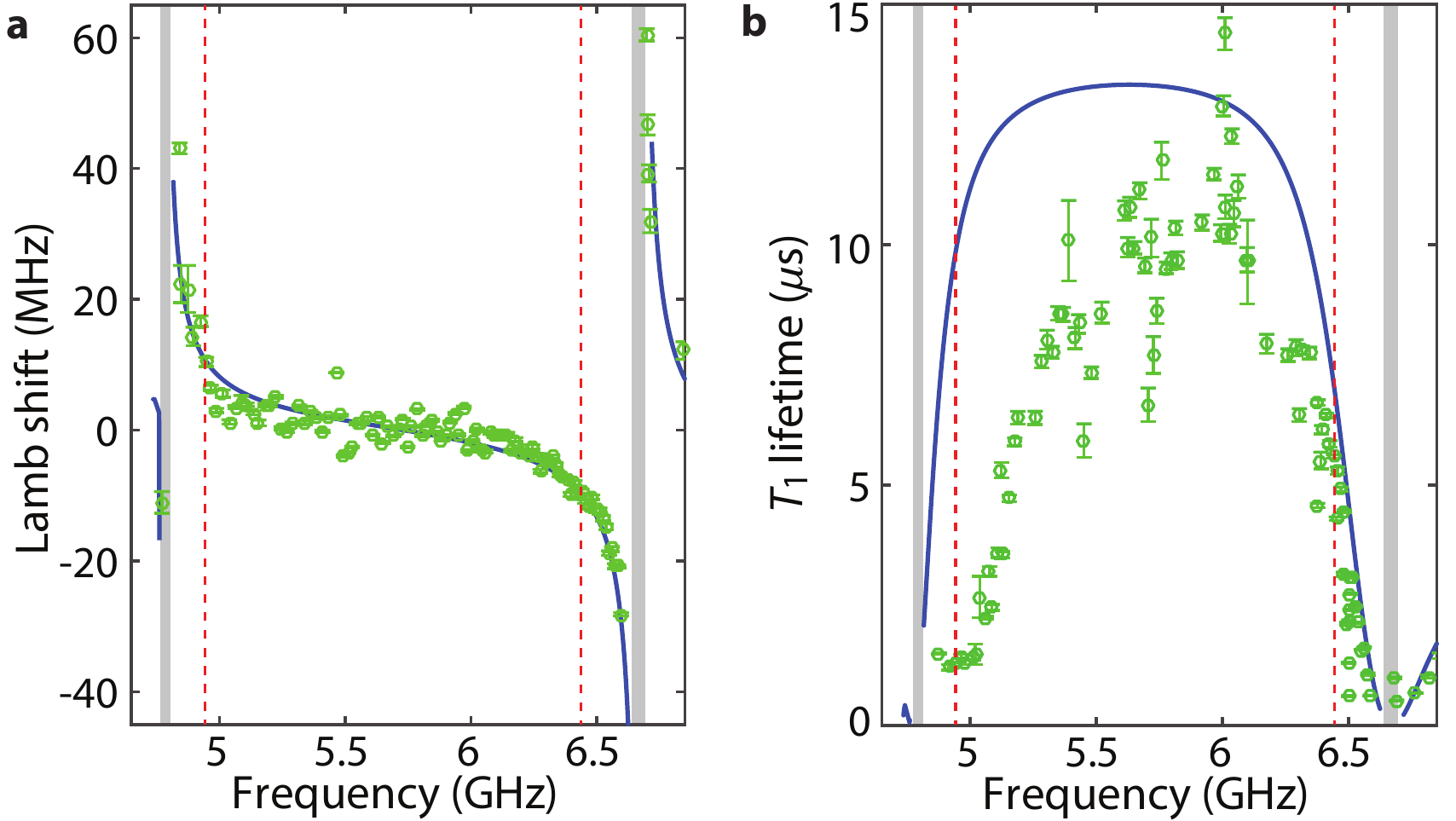}}
\caption{\textbf{Measured dispersive and dissipative qubit dynamics.} \textbf{a}, Qubit frequency Lamb shift versus frequency.  \textbf{b}, Qubit lifetime versus frequency.  The open circles show experimental data and the solid line presents a theory fit. The dashed red lines mark the estimated position of the band-edge corresponding to an infinite-length structure and the shaded grey regions correspond to anti-crossing with the first individual resonances near the band-edges of the finite structure. For calculating the Lamb shift, the bare qubit frequency is calculated as a function of flux bias $\Phi$ as $\hbar \omega_{ge} = \sqrt{8 E_C E_J(\Phi)} - E_C$ using the extracted values of $E_C$, $E_J$, and assuming the symmetrical SQUID flux bias relation $E_J(\Phi) = E_{J,\mathrm{max}} \cos(2\pi\Phi/\Phi_0)$. The lifetime characterization is performed in the time domain where the qubit is initially excited with a $\pi$ pulse through the XY drive. The excited state population, determined from the state-dependent dispersive shift of a close-by band-edge waveguide mode, is measured subsequent to a delay time during which the qubit freely decays.} 
\label{fig:ExpBandGap}
\vspace{-0.5 cm}
\end{figure}

To further probe the electromagnetic properties of the metamaterial waveguide we couple it to a superconducting qubit. In this work we use a transmon qubit \cite{Koch:2007gz, Barends:2013kz} with the fundamental resonance frequency $\nu_{ge} = 7.9$ GHz and Josephson energy to single electron charging energy ratio of $E_J/E_C \approx 100$ at zero flux bias (details of our qubit fabrication methods can also be found in Ref.~\cite{Keller2017}). Figure\,\ref{fig:QBWGSEM}b shows the geometry of the device where the qubit is capacitively coupled to one end of the waveguide and the other end is capacitively coupled to a $50$-$\Omega$ CPW transmission line. This geometry allows for forming narrow individual modes in the transmission band of the metamaterial, which can be used for dispersive qubit state read-out \cite{Wallraff:2005in} from reflection measurements at the $50$-$\Omega$ CPW input port (see Fig.\,\ref{fig:QBWGSEM}b). Within the bandgap the qubit is self-dressed by virtual photons which are emitted and re-absorbed due to the lack of escape channels for the radiation.  Near the band-edges surrounding the bandgap, where the LDOS is rapidly varying with frequency, this can result in a large anomalous Lamb shift of the dressed qubit frequency \cite{John:1991es,Kofman:1994gi}. To observe this effect, we tune the qubit frequency using a flux bias \cite{Barends:2013kz} and find the frequency shift by subtracting the measured frequency from the expected frequency of the qubit as a function of flux bias. Figure\,\ref{fig:ExpBandGap}a shows the measured frequency shift as a function of tuning. It is evident that the qubit frequency is repelled from the band-edges on the two sides, as a result of the asymmetric density of states near the cut-off frequencies. The measured frequency shift is approximately $10$~MHz at the band-edges ($0.2\%$ of the qubit frequency), in excellent agreement with the circuit theory model (see App.~\ref{App:E}).
 
\begin{figure}[t!]
\centerline{\includegraphics[width = \columnwidth]{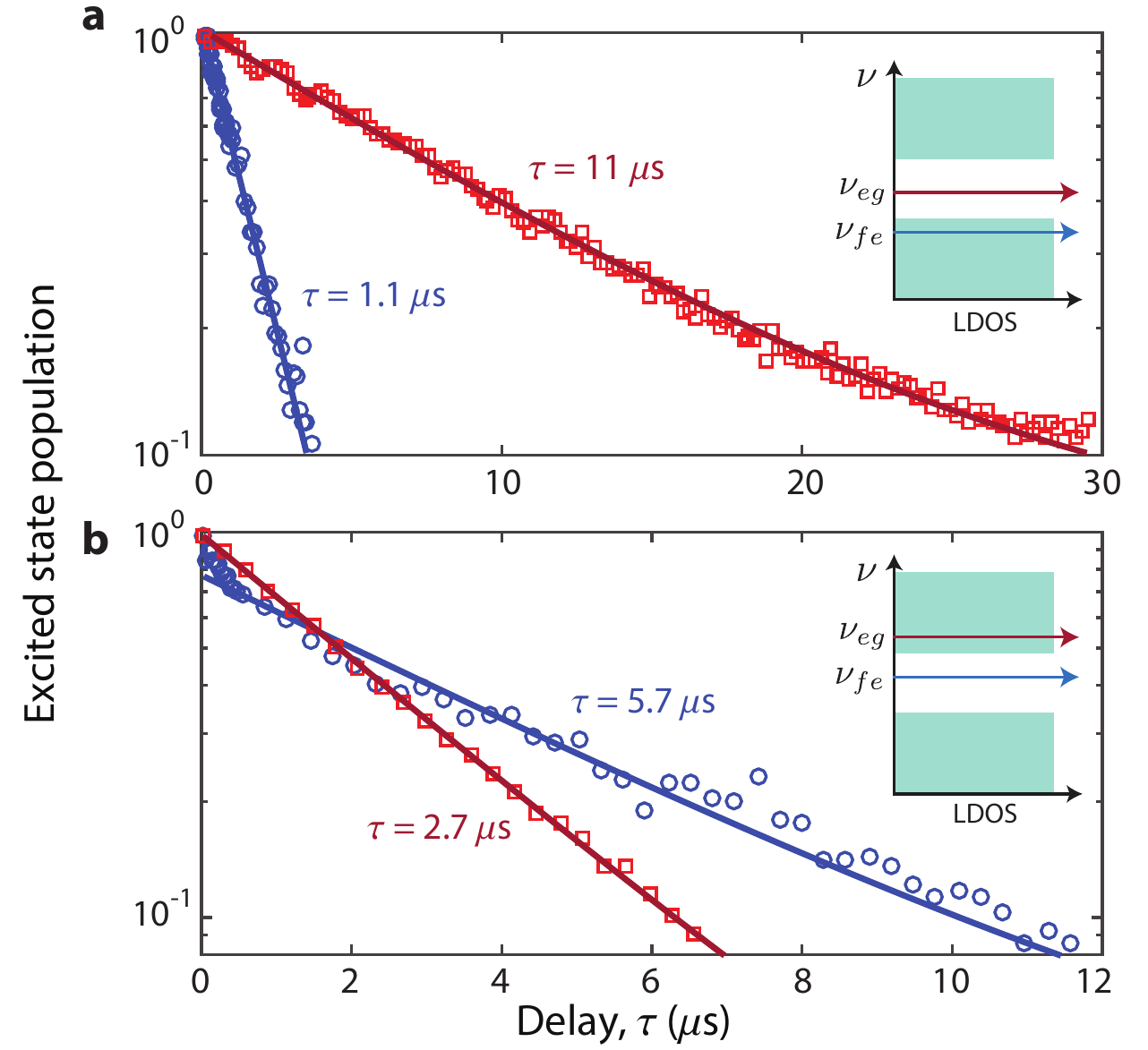}}
\caption{\textbf{State-selective enhancement and inhibition of radiative decay.} \textbf{a}, Measurement with the $g$-$e$ transition tuned into the bandgap, with the $f$-$e$ transition in the lower transmission band. \textbf{b}, Measurement with the $g$-$e$ transition tuned into the upper transmission band, with the $f$-$e$ transition in the bandgap. For measuring the $f$-$e$ lifetime, we initially excite the third energy level $\ket{f}$ via a two-photon $\pi$ pulse at the frequency of $\omega_{gf} /2$. Following the population decay in a selected time interval, the population in  $\ket{f}$ is mapped to the ground state using a second $\pi$ pulse. Finally the ground state population is read using the dispersive shift of a close-by band-edge resonance of the waveguide. $g$-$e$ ($f$-$e$) transition data shown as red squares (blue circles)} 
\label{fig:ThreeLevels}
\end{figure}

Another signature of the qubit-waveguide interaction is the change in the rate of spontaneous emission of the qubit. Tuning the qubit into the bandgap changes the localization length of the waveguide photonic state that dresses the qubit. Since the finite waveguide is connected to an external port which acts as a dissipative environment, the change in localization length $\ell(\omega)$ is accompanied by a change in the radiative lifetime of the qubit $T_\text{rad}(\omega)\propto e^{2x/\ell(\omega)}$, where $x$ is the total length of the waveguide.  Figure\,\ref{fig:ExpBandGap}b shows the measured qubit lifetime ($T_1$) as a function of its frequency in the bandgap.  It is evident that the qubit lifetime drastically increases inside the bandgap, where spontaneous emission into the output port is greatly suppressed due to the reduced localization length of the photon bound state.  Deep within the bandgap one observes the appearance of multiple narrow spectral features in the measured frequency dependence of the qubit lifetime.  These features, attributable to parasitic ``box'' modes of our chip packaging, highlight the ability of the metamaterial waveguide to enable effectively-dissipation-free probing of the qubit's environment over a broad spectral range ($> 1$~GHz).  As the qubit frequency approaches the band-edges, the lifetime is sharply reduced because of the increase in the localization length of the waveguide modes. The slope of the lifetime curve at the band-edge can be shown to be directly proportional to the group delay, $\left|{\partial  T_\text{rad}}/{\partial  \omega}\right| = T_\text{rad} \tau_\text{delay}$ (see App.~\ref{App:E}). We observe a $24$-fold enhancement in the lifetime of the qubit near the upper band-edge, corresponding to a maximum group index of $n_g=450$ right at the band-edge.
 
In addition to radiative decay into the output channel, losses in the resonators in the waveguide also contribute to the qubit's excited state decay. Using a low power probe in the single-photon regime we have measured intrinsic $Q$-factors of $7.2 \pm 0.4 \times 10^4$ for the individual waveguide modes between $4.6$-$7.4$~GHz. The solid line in Fig.\,\ref{fig:ExpBandGap}b shows a fitted theoretical curve which takes into account the loss in the waveguide along with a phenomenological intrinsic lifetime of the qubit. While the measured lifetime near the upper band is in excellent agreement with the theoretical model, the data near the lower band shows significant departure from the model. We attribute this departure in the lower band to the presence of a spurious resonance or resonances near the lower band-edge. Possible candidates for such spurious modes include the asymmetric ``slotline'' modes of the metamaterial waveguide, which are weakly coupled to our symmetrically grounded CPW line but may couple to the qubit. Further study of the spectrum of these modes and possible methods for suppressing them using cross-over connections \cite{Chen:2014he} will be a topic of future studies.
 
The sharp variation in the photonic LDOS near the metamaterial waveguide band-edges may also be used to engineer the multi-level dynamics of the qubit. A transmon qubit, by construct, is a nonlinear quantum oscillator and thus it has a multilevel energy spectrum. In particular, a third energy level ($\ket{f}$) exists at the frequency $\omega_{gf} = 2\omega_{ge} - E_C/\hbar$. Although the transition $g$-$f$ is not allowed because of the selection rules, the $f$-$e$ transition is allowed and has a dipole moment that is $\sqrt{2}$ larger than the fundamental transition \cite{Koch:2007gz}. This is reminiscent of the scaling of transition amplitudes in a harmonic oscillator and results in a second transition lifetime that is half of the fundamental transition lifetime for a uniform density of states in the electromagnetic bath. Nonetheless, the sharply varying density of states in the metamaterial can lead to strong suppression or enhancement of the spontaneous emission for each transition. Figure\,\ref{fig:ThreeLevels} shows the measured lifetimes of the two transitions for two different spectral configurations. In the first scenario, we enhance the ratio of the lifetimes $T_{eg}/T_{fe}$ by situating the fundamental transition frequency inside in the bandgap while having the second transition positioned inside the lower transmission band. The situation is reversed in the second configuration, where the fundamental frequency is tuned to be within the upper energy band while the second transition lies inside the gap. In our fabricated qubit, the second transition is $290$~MHz lower than the fundamental transition frequency at zero flux bias, which allows for achieving large lifetime contrast in both configurations.

Compact, low loss, low disorder superconducting metamaterials, as presented here, can help realize more scalable superconducting quantum circuits with higher levels of complexity and functionality in several regards.  They offer a method for densely packing qubits -- both in spatial and frequency dimensions -- with isolation from the environment by operation in forbidden bandgaps, and yet with controllable connectivity achieved via bound qubit-waveguide polaritons~\cite{Douglas:2015hd,Calajo:2016fz}.  Moreover, the ability to selectively modify the transition lifetimes provides simultaneous access to long-lived metastable qubit states as well as short-lived states strongly coupled to waveguide modes.  This approach realizes an effective $\Lambda$-type level structure for the transmon, and can be used to create state-dependent bound state localization lengths, quantum nonlinear media for propagating microwave photons~\cite{Nikoghosyan2010,Douglas:2016kr,Albrecht2017}, or as recently demonstrated, to realize spin-photon entanglement and high-bandwidth itinerant single microwave photon detection~\cite{Inomata:2016jc, Besse:2017wh}.  Combined, these attributes provide a unique platform for studying the many-body physics of quantum photonic matter~\cite{Greentree:2006jg,Hartmann:2006kv,Houck2012,Noh2016}. 

\section*{Acknowledgments}

We would like to thank Paul Dieterle, Ana Asenjo Garcia and Darrick Chang for fruitful discussions regarding waveguide QED.  This work was supported by the AFOSR MURI Quantum Photo4nic Matter (grant 16RT0696), the AFOSR MURI Wiring Quantum Networks with Mechanical Transducers (grant FA9550-15-1-0015), the Institute for Quantum Information and Matter, an NSF Physics Frontiers Center (grant PHY-1125565) with support of the Gordon and Betty Moore Foundation, and the Kavli Nanoscience Institute at Caltech.  M.M. (A.J.K., A.S.) gratefully acknowledges support from a KNI (IQIM) Postdoctoral Fellowship.



%

\appendix

\section{Band structure calculation}
\label{App:A}

\subsection{Quantization of a periodic resonator-loaded waveguide}

\begin{figure*}[t]
\begin{center}
\ctikzset{bipoles/capacitor/height/.initial=.40}   
\ctikzset{bipoles/capacitor/width/.initial=.20}
\ctikzset{bipoles/generic/width/.initial=.25}
\ctikzset{bipoles/generic/height/.initial=.25}
\begin{circuitikz}[scale=0.6] 
 \draw (-11,1) node[anchor=east] {$\cdots $}
  to [short, -*] (-11,1) node[anchor=south] {$\Phi^a_{n-1}$}
  to [cute inductor, *-, l_=$L_0$] (-8,1) node[anchor=south] {}
  to [short, -*] (-5,1) node[anchor=south] {$\Phi^a_{n}$}
  to [cute inductor, *-, l_=$L_0$] (0,1) node[anchor=south] {}
       to [short,-*] (3,1) node[anchor=south] {$\Phi^a_{n+1}$}
    to [short] (5, 1) node[anchor=west] {$\cdots$};
 \draw (-1, -3) node[ground] {}
    to [capacitor, l_=$C_0$] (-1, 1) {};
 \draw (2, 1) {}
    to [capacitor, -*, l_=$C_k$] (2, -0.5) node[anchor=north] {$\Phi^b_{n+1}$};
  \draw (3, -0.5) {}
    to [capacitor, l_=$C_r$] (3, -3) {};
 \draw (1, -0.5) {}
    to [cute inductor, l_=$L_r$]  (1, -3) {};
 \draw (1, -0.5) {}
   to [short] (3, -0.5) {};

   \draw (-5, 1) {}
    to [capacitor, -*, l_=$C_k$] (-5, -0.5) node[anchor=north] {$\Phi^b_{n}$};
  \draw (-4, -0.5) {}
    to [capacitor, l_=$C_r$] (-4, -3) {};
 \draw (-6, -0.5) {}
    to [cute inductor, l_=$L_r$]  (-6, -3) {};
 \draw (-6, -0.5) {}
   to [short] (-4, -0.5) {};
    \draw (-8, -3) node[ground] {}
    to [capacitor, l_=$C_0$] (-8, 1) {};
 \draw (-11, -3) node[anchor=east] {$\cdots$}
    to (5, -3) node[anchor=west] {$\cdots$};


 \draw (-11,-6) node[anchor=east] {$\cdots $}
  to [short] (-11,-6) node[anchor=south] {}
  to [cute inductor, l_=$L_0$] (-8,-6) node[anchor=south] {};
  \draw (-8,-5) {}
    to [cute inductor, l_=$L_r$] (-6,-5) node[anchor=south] {}
    to  [capacitor, l_=$C_r$] (-4, -5) {} node[anchor=south] {};
 \draw (-8,-7) {}
    to [cute inductor, l_=$L_k$] (-4, -7) {} node[anchor=south] {};
      \draw (-8,-5) {}
     to [short] (-8,-7){};
           \draw (-4,-5) {}
     to [short] (-4,-7){};
 \draw (-4,-6) {}
    to [short] (-3, -6) {};

 \draw (-3, -10) node[ground] {}
    to [capacitor, l_=$C_0$] (-3, -6) {};

  \draw (-3,-6) {}
  to [cute inductor, l_=$L_0$] (0,-6) node[anchor=south] {};
  \draw (0,-5) {}
    to [cute inductor, l_=$L_r$] (2,-5) node[anchor=south] {}
    to  [capacitor, l_=$C_r$] (4, -5) {} node[anchor=south] {};
 \draw (0,-7) {}
    to [cute inductor, l_=$L_k$] (4, -7) {} node[anchor=south] {};
      \draw (0,-5) {}
     to [short] (0,-7){};
           \draw (4,-5) {}
     to [short] (4,-7){};
 \draw (4,-6) {}
    to [short] (4.5, -6) node[anchor=west] {$\cdots$};

    \draw (-11, -10) node[ground] {}
    to [capacitor, l_=$C_0$] (-11, -6) {};
 \draw (-11, -10) node[anchor=east] {$\cdots$}
    to (4.5, -10) node[anchor=west] {$\cdots$};

\end{circuitikz}
\end{center}
\caption{Circuit diagram of metamaterial waveguides made from periodic arrays of transmission line sections loaded with capacitively coupled resonators (top), and inductively loaded resonators (bottom). }\label{fig:setup-transmission-line}
\end{figure*}
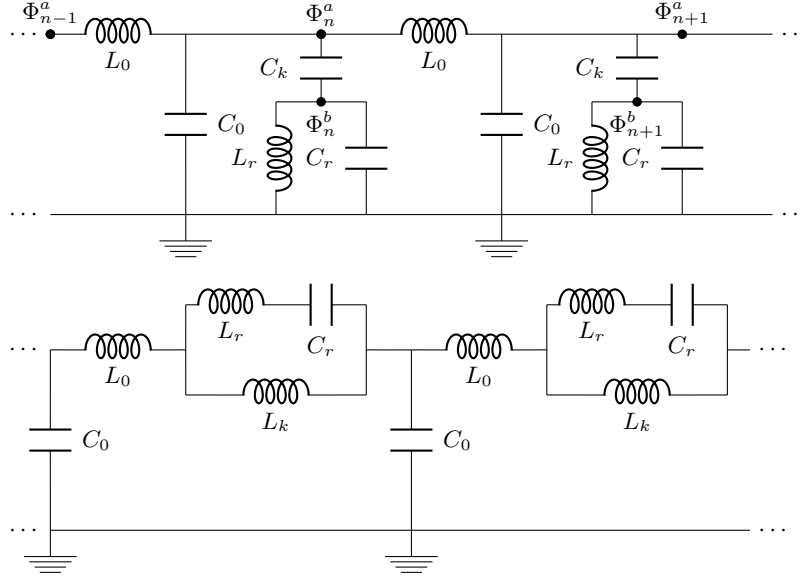

We consider the case of a waveguide that is periodically loaded with microwave resonators. Fig.\,\ref{fig:setup-transmission-line}  depicts a unit cell for this configuration. The Lagrangian for this system can be readily written as \cite{Devoret:1995vn}
\begin{align}\label{eq:hamiltonian1}
{L} =& \sum_n \bigg[      \frac{1}{2} C_0  {[\dot{\Phi}^a_n]}^2   - \frac{[{{\Phi}^a_n-{\Phi}^a_{n-1}}]^2}{2 L_0}  \nonumber\\
& + \frac{1}{2} C_r  {[\dot{\Phi}^b_n]}^2  +  \frac{1}{2} C_k {[\dot{\Phi}^a_n-\dot{\Phi}^b_n]}^2 -  \frac{{[{\Phi}^b_n]}^2}{2L_r}  \bigg].
\end{align}
In order to find solutions in form of traveling waves, it is easier to work with the Fourier transform of node fluxes. We use the following convention for defining the (discrete) Fourier transformation
\begin{align}
&{\Phi}^{a,b}_\kappa = \frac{1}{{\sqrt{M}}} \sum_{n=-N}^N e^{-i2\pi(\kappa/M)n}\Phi^{a,b}_n ,
\end{align}
where $M = 2N+1$ is the total number of periods in the waveguide. 
 Using the Fourier relation we find the Lagrangian in $k$-space as
\begin{align}
{L} =& \sum_\kappa \bigg[      \frac{1}{2}( C_0 +C_k ) {|\dot{\Phi}^a_\kappa|}^2 -  \left|1- e^{-i2\pi(\kappa/M)}\right|^2\frac{{|{\Phi}^a_\kappa|}^2}{2L_0} \nonumber\\
& \frac{1}{2}( C_k +C_r ) {|\dot{\Phi}^b_\kappa|}^2 - \frac{{|{\Phi}^b_\kappa|}^2}{2L_r} -C_k \frac{\dot{\Phi}^b_\kappa\dot{\Phi}^a_{-\kappa} +\dot{\Phi}^b_{-\kappa}\dot{\Phi}^a_{\kappa}}{2} \bigg].
\end{align}
To proceed further, we need to find the canonical node charges  which are defined as $Q^{a,b}_\kappa = \frac{\partial L}{\partial \dot{\Phi}^{a,b}_\kappa} $, and subsequently derive the Hamiltonian of the system by using a Legendre transformation. Doing so we find
\begin{align}
{H} =& \sum_\kappa \bigg[ \frac{{Q^a_\kappa}{Q^a_{-\kappa}}}{2C_0'} + \left|1- e^{-i2\pi(\kappa/M)}\right|^2\frac{{{\Phi}^a_\kappa}{{\Phi}^a_{-\kappa}}}{2L_0}+ \nonumber\\
&\frac{{Q^b_\kappa}{Q^b_{-\kappa}}}{2 C_r'} + \frac{{{\Phi}^b_\kappa}{{\Phi}^b_{-\kappa}}}{2L_r} +\frac{{Q^a_\kappa}{Q^b_{-\kappa}} +{Q^a_{-\kappa}}{Q^b_{\kappa}}}{2 C_k' }  \bigg].
\end{align}
Here, we have defined the following quantities
\begin{align}
C_0' = \frac{C_kC_r + C_kC_0 +C_0C_r  }{C_k +C_r},\\
C_r' = \frac{C_kC_r + C_kC_0 +C_0C_r  }{C_k +C_0},\\
C_k' = \frac{C_kC_r + C_kC_0 +C_0C_r  }{C_k }.
\end{align}
The canonical commutation relation \mbox{$[{\Phi}^i_{\kappa},Q^j_{-\kappa'}] = i\hbar \delta_{i,j}\delta_{\kappa,\kappa'}$} allows us to define the following annihilation operators as a function of charge and flux operators
\begin{align}
\hat{a}_\kappa = &\sqrt{\frac{C_0' \Omega_k}{2\hbar}} \left( {\Phi}^a_{\kappa} +\frac{i}{C_0' \Omega_k} Q^a_{\kappa}  \right),\\
\hat{b}_\kappa = &\sqrt{\frac{C_r' \omega_0}{2\hbar}} \left( {\Phi}^b_{\kappa} +\frac{i}{C_r' \omega_0} Q^b_{\kappa}  \right).
\end{align}
Here, we have defined the resonance frequency for each mode as 
\begin{align}
&\Omega_k = \sqrt{\frac{4 {\mathrm{sin}}^2(kd/2)}{L_0 C_0'}},\\
&\omega_0 = \frac{1}{\sqrt{L_r C_r'}},
\end{align}
where $k = (2\pi\kappa)/(Md)$ is the wavenumber. It is evident that $\Omega_k $ has the expected dispersion relation of a discrete periodic transmission line and $\omega_0 $ is the resonance frequency of the loaded microwave resonators. Using the above definitions for $\hat{a}_\kappa, \hat{b}_\kappa$
\begin{align}\label{Eq:HamiltonianExact}
\hat{H} &= \frac{\hbar}{2} \sum_{k } \bigg[   \Omega_k \left( \hat{a}_k^\dagger\hat{a}_k  + \hat{a}_{-k}\hat{a}_{-k}^\dagger \right)+  {\omega_0} \left( \hat{b}^\dagger_k\hat{b}_k + \hat{b}_{-k}\hat{b}_{-k}^\dagger \right)\nonumber \\
& - g_k \left(  \hat{b}_{-k} -   \hat{b}_{k}^\dagger \right) \left( \hat{a}_{k} -   \hat{a}_{-k}^\dagger \right) - g_k    \left( \hat{a}^\dagger_{k} -   \hat{a}_{-k} \right)\left(  \hat{b}^\dagger_{-k} -   \hat{b}_{k} \right) \bigg],
\end{align}
along with the coupling coefficient 
\begin{align}
g_k = \frac{\sqrt{{C_0' C_r'}}}{2 C_k'}  \sqrt{{\omega_0 \Omega_k}} = \frac{C_k  \sqrt{{\omega_0 \Omega_k}}}{2\sqrt{{(C_0+C_k) (C_r+C_k)}}} .
\end{align}
An alternative structure for coupling microwave resonators is depicted in the bottom panel of Fig.\,\ref{fig:setup-transmission-line}. In this geometry, the coupling is controlled by the inductive element $L_k$. Repeating the analysis above for this case, we find

\begin{align}
&\Omega_k = \sqrt{\frac{4 {\mathrm{sin}}^2(kd/2)}{C_0 L_0'}},\\
&\omega_0 = \frac{1}{\sqrt{C_r L_r'}},\\
&g_k = \frac{\sqrt{{L_0' L_r'}}}{2 L_k'}  \sqrt{{\omega_0 \Omega_k}}.
\end{align}
We have defined the modified inductance values as
\begin{align}
L_0' = \frac{L_kL_r + L_kL_0 +L_0L_r  }{L_k +L_r},\\
L_r' = \frac{L_kL_r + L_kL_0 +L_0L_r  }{L_k +L_0},\\
L_k' = \frac{L_kL_r + L_kL_0 +L_0L_r  }{L_k }.
\end{align}

\subsection{Band structure calculation with RWA}
Using the rotating wave approximation, the Hamiltonian in Eq.\,(\ref{Eq:HamiltonianExact}) can be simplified to
\begin{align}
\hat{H} &= {\hbar} \sum_{k } \bigg[   \Omega_k  \hat{a}_k^\dagger\hat{a}_k  +  {\omega_0}  \hat{b}^\dagger_k\hat{b}_k + g_k \left( \hat{b}_{k}^\dagger  \hat{a}_{k} +  \hat{a}^\dagger_{k}   \hat{b}_{k} \right) \bigg],
\end{align}
The simplified Hamiltonian can be written in the compact form 
\begin{align}
\hat{H} = \hbar \sum_{k } {\mathbf{x}}_k^\dagger \mathbf{H}_k  {\mathbf{x}}_k, 
\end{align}
where 
\begin{align}
 \mathbf{H}_k= 
  \begin{bmatrix}
    \Omega_k &  g_k \\
    g_k & \omega_0 \\
  \end{bmatrix}
  ,
  {\mathbf{x}}_k= 
    \begin{bmatrix}
   &\hat{a}_{k}   \\
   &\hat{b}_{k} 
  \end{bmatrix}.
\end {align}
We desire to transform the Hamiltonian to a diagonalized form
  \begin{align}
   \mathbf{\tilde{H}}_k= 
  \begin{bmatrix}
    \omega_{+,k} & 0 \\
     0 & \omega_{-,k}
  \end{bmatrix}.
  \end {align}
It is straightforward to use the eigenvalue decomposition to find $\omega_{\pm,k}$ as
 \begin{align}
\omega_{\pm,k} = \frac{1}{2} \left[ \left(\Omega_k + \omega_0 \right) \pm \sqrt{{\left( \Omega_k - \omega_0 \right)}^2 + 4  g_k^2} \right],
\end {align}
along with the corresponding eigenstates
 \begin{align}
\hat{\alpha}_{\pm,k}  = \frac{(\omega_{\pm,k} - \omega_0)}{\sqrt{ {(\omega_{\pm,k} - \omega_0)}^2 + g_k^2}}\hat{a}_{k} +  \frac{g_k}{\sqrt{ {(\omega_{\pm,k} - \omega_0)}^2 + g_k^2}}\hat{b}_{k} .
\end {align}
\subsection{Band structure calculation beyond RWA}
The exact Hamiltonian in Eq.\,(\ref{Eq:HamiltonianExact}) can be written in the compact form 
\begin{align}
\hat{H} = \frac{\hbar}{2} \sum_{k } {\mathbf{x}}_k^\dagger \mathbf{H}_k  {\mathbf{x}}_k, 
\end{align}
where 
\begin{align}
 \mathbf{H}_k= 
  \begin{bmatrix}
    \Omega_k & 0 & g_k & -g_k \\
    0 & \Omega_k & -g_k & g_k \\
    g_k & -g_k & \omega_0 & 0 \\
    -g_k & g_k & 0 & \omega_0
  \end{bmatrix}
  ,
  {\mathbf{x}}_k= 
    \begin{bmatrix}
   &\hat{a}_{k}   \\
   &\hat{a}_{-k}^\dagger  \\
   &\hat{b}_{k} \\
    &\hat{b}_{-k}^\dagger 
  \end{bmatrix}.
\end {align}
 To find the eigenstates of the system, we can use a linear transform to map the state vector ${\mathbf{\tilde{x}}}_k = \mathbf{S}_k{\mathbf{{x}}}_k$ such that ${\mathbf{x}}_k^\dagger \mathbf{H}_k  {\mathbf{x}}_k =  {\mathbf{\tilde{x}}}_k^\dagger \mathbf{\tilde{H}}_k  {\mathbf{\tilde{x}}}_k$ with the transformed diagonal Hamiltonian matrix
 \begin{align}
 \mathbf{\tilde{H}}_k= 
  \begin{bmatrix}
    \omega_{+,k} & 0 & 0 & 0 \\
    0 & \omega_{+,k} & 0 & 0 \\
    0 & 0 & \omega_{-,k} & 0 \\
    0 & 0 & 0 & \omega_{-,k}.
  \end{bmatrix}
\end {align}
In order to preserve the canonical commutation relations, the matrix $\mathbf{S}_k$ has to be symplectic, i. e. $\mathbf{S}_k = \mathbf{S}_k \mathbf{J} \mathbf{S}_k^\dagger$, with the matrix $\mathbf{J}$ defined as
 \begin{align}
\mathbf{J}= 
  \begin{bmatrix}
    1 & 0 & 0 & 0 \\
    0 & -1 & 0 & 0 \\
    0 & 0 & 1 & 0 \\
    0 & 0 & 0 & -1.
  \end{bmatrix}
\end {align}
A linear transformation (such as $\mathbf{S}_k$) that diagonalizes a set of quadratically coupled boson fields while preserving their canonical commutation relations is often referred to as a Bogoliubov-Valatin transformation. While it is generally difficult to find the transform matrix $\mathbf{S}_k$, it is easy to find the eigenvalues of the diagonalized Hamiltonian by exploiting some of the properties of $\mathbf{S}_k$. Note that since $\mathbf{S}_k = \mathbf{S}_k \mathbf{J} \mathbf{S}_k^\dagger$, the matrices $\mathbf{J}\mathbf{\tilde{H}}_k$ and $\mathbf{J}\mathbf{{H}}_k$ share the same set of eigenvalues. The eigenvalues of $\mathbf{J}\mathbf{\tilde{H}}_k$ are the two frequencies $\omega_{\pm,k}$, and thus we have
 \begin{align}\label{Eq:ExactDisp}
\omega_{\pm,k}^2 = \frac{1}{2} \left[ \left(\Omega_k^2 + \omega_0^2 \right) \pm \sqrt{{\left( \Omega_k^2 - \omega_0^2 \right)}^2 + 16 \omega_0 \Omega_k g_k^2} \right].
\end {align}

\subsection{Circuit theory derivation of the band structure}

Consider the pair of equations that describe the propagation of a monochromatic electromagnetic wave of the form $v(x,t) =V(x) e^{-ikx}e^{i\omega t}$ (along with the corresponding current relation) inside a transmission line
\begin{align} \label{Eq:Telegrapher}
\frac{\mathrm d}{\mathrm d x}  V(x)& = - Z(\omega) I(x), \nonumber \\
\frac{\mathrm d }{\mathrm d x} I(x)& = - Y(\omega) V(x).
\end{align} 
Here, $Z(\omega)$ and $Y(\omega)$ are frequency dependent impedance and admittance functions that model the linear response of the series and parallel portions of a transmission line with length $d$. It is straightforward to check that the solutions to these equation satisfy $k(\omega) = n \omega/c = \sqrt{-Z(\omega)Y(\omega)}/d$. For a loss-less waveguide and in the absence of dispersion we have $Z(\omega) = i \omega L_0$ and $Y(\omega) = i \omega C_0$, and thus we find the familiar dispersion relation $k(\omega) = \omega\sqrt{L_0 C_0}/d$. Nevertheless, the pair of equations above remain valid for arbitrary impedance and admittance functions  $Z(\omega)$ and $Y(\omega)$, provided that the dimension of the model circuit remains much smaller than the wavelength under consideration. In this model, a real and negative quantity for the product $ZY$ results in an imaginary wavenumber and subsequently creates a stop band in the dispersion relation. This situation can be achieved by periodically loading a transmission line with an array of resonators \cite{Karyamapudi:iz, OBrien:2014fc}. Assuming a unit length of $d$ we find
\begin{align} \label{Eq:ResDisp}
k^2 =  {\left(\frac{\omega}{c}\right)}^2 {n}^2 \left[1+ \frac{2c \gamma_e}{n d}  \frac{1}{\omega_0^2 - \omega^2} \right] .
\end{align} 
Here, $\omega_0$ is the resonance frequency, and $\gamma_e$ is the external coupling decay rate of an individual resonator in the array. For moderate values of gap-midgap ratio ($\Delta/\omega_m$), the frequency gap can be found as
\begin{align}
\Delta = \frac{c }{n d} \left(\frac{\gamma_e}{\omega_0}\right),
\end{align} 
and $\omega_m = \omega_0 + \Delta/2$. We have defined the gap as the range of frequencies where the wavenumber is imaginary.

Although a microwave resonator can be realized by using a two-elements $LC$-circuit, the three-element circuits in Fig.\,\ref{fig:setup-transmission-line} provide an additional degree of freedom which enables setting the coupling $\gamma_e$ independent of the resonance frequency $\omega_0$. Using circuit theory, it is straightforward to show 
\begin{align}
\omega_0 = \frac{1}{\sqrt{L_r(C_r+ C_k)}},\\
\gamma_e =\frac{Z_0}{2L_r} {\left(\frac{C_k}{C_r+ C_k}\right)}^2 .
\end{align} 
Here, $Z_0$ is the characteristic impedance of the unloaded waveguide. It is easy to check that for small values of $C_k/C_r$, the resonance frequency is only a weak function of $C_k$. As a result, it is possible to adjust the coupling rate $\gamma_e$ by setting the capacitor $C_k$ while keeping the resonance frequency almost constant. Figure\,\ref{fig:setup-transmission-line} also depicts an alternative strategy for coupling microwave resonators to the waveguide. In this design, the inductive element $L_k$ is used to set the coupling in a ``current divider" geometry. We provide experimental results for implementation of bandgap waveguide based on both designs in the next section.

While the ``continuum" model described above provides a heuristic explanation for formation of bandgap in a waveguide loaded with resonators, its results remains valid as far as $k \ll 2\pi/d$. To avoid this approximation, we can use the transfer matrix method to find the exact dispersion relation for a system with discrete periodic symmetry \cite{Pozar:1998wp}. In this case Eq.\,(\ref{Eq:ResDisp}) is modified to
\begin{align}\label{Eq:ResDisp_discrete}
  \cos{(kd)}  = 1-{\left(\frac{\omega}{c}\right)}^2 \frac{{n}^2 d^2}{2} -  \frac{ n d \gamma_e}{c}  \frac{\omega^2}{\omega_0^2 - \omega^2}  .
\end{align}
Note that this relation still requires $d$ to be much smaller than the wavelength of the unloaded waveguide $\lambda = 2\pi c/ (n\omega)$. 

\subsection{Dispersion and group index near the band-edges}

Equation (\ref{Eq:ExactDisp}) can be reversed to find the wavenumber $k$ as a function of frequency. Assuming, a linear dispersion relation of the form $k= n\Omega_k /c$ for the bare waveguide we find
\begin{align}\label{Eq:ExactDisp2}
k = \frac{n\omega}{c} \sqrt{\frac{\omega^2-\omega_{c+}^2}{\omega^2-\omega_{c-}^2}}.
\end{align}
Here, $\omega_{c+} = \omega_0$ and $\omega_{c-} = \omega_0 \sqrt{1- 4g_k^2/(\Omega_k\omega_0)} $ are the upper and lower cut-off frequencies, respectively. The quantity $g_k^2/(\Omega_k\omega_0)$ is a unit-less parameter quantifying the size of the bandgap and is independent of the wavenumber $k$. 

The dispersion relation can be written in simpler forms by expanding the wavenumber in the vicinity of the two band-edges
\begin{align}
k =
  \begin{cases}
    \frac{n\omega_{c-}}{c} \sqrt{ \frac{\Delta}{-\delta_{-}} }        & \quad \text{for } \omega\approx \omega_{c-}, \\
    \frac{n\omega_{c+}}{c} \sqrt{ \frac{\delta_{+}}{\Delta} }    & \quad \text{for }  \omega\approx \omega_{c+}.
  \end{cases}
\end{align}
Here, $\Delta = \omega_{c+}-\omega_{c-}$ is the frequency span of the bandgap and $\delta_\pm = \omega- \omega_{c\pm}$ are the detunings from the band-edges.

The form of the dispersion relation Eq.\,(\ref{Eq:ExactDisp}) suggests that the maxima of the group index happens near the band-edges. Having the wavenumber, we can readily evaluate the group velocity $v_g = {\partial  \omega }/{\partial  k}$ and find the group index $n_g = c/v_g$ as
\begin{align}\label{Eq:suppGroupIndx}
n_g =
  \begin{cases}
    {\frac{{n\omega_{c-}} \sqrt{\Delta}}{\sqrt{-4{(\delta_{-} - i\gamma_i)}^3}}  }        & \quad \text{for } \omega\approx \omega_{c-}, \\
     { \frac{ {n\omega_{c+}} }{\sqrt{4\Delta (\delta_{+} - i\gamma_i)}} }    & \quad \text{for }  \omega\approx \omega_{c+}.
  \end{cases}
\end{align}
Note that we have replaced $\delta_\pm$ with $\delta_\pm - i\gamma_i$ to account for finite internal quality factor of the resonators in the structure.

\section{Coupling a Josephson junction qubit to a metamaterial waveguide}
\label{App:B}

We consider the coupling of  a Josephson junction qubit to the metamaterial waveguide. Assuming rotating wave approximation, the Hamiltonian of this system can be written as
\begin{align}
\hat{H} &= {\hbar} \sum_{k } \bigg[   \omega_k  \hat{a}_k^\dagger\hat{a}_k  +  \frac{\omega_q}{2}  \hat{\sigma}_z + f_k \left( \hat{a}_{k}^\dagger  \hat{\sigma}^{-} +   \hat{a}_{k}  \hat{\sigma}^+ \right).\bigg]
\end{align}
Here $ f_k$ is the coupling factor of the qubit to the waveguide photons, and $ \omega_k = \omega_{\pm,k} $, where the plus or minus sign is chosen such that the qubit frequency $\omega_q$ lies within the band. Without loss of generality, we assume $ f_k$ to be a real number. The Heisenberg equations of motions for the qubit and the photon operators can be written as
\begin{align}
&\frac{\partial }{\partial  t} \hat{a}_k = -i \omega_k \hat{a}_k - i f_k  \hat{\sigma}^- \\
&\frac{\partial }{\partial  t}  \hat{\sigma}^- = -i \omega_q \hat{\sigma}^- - i  \sum_k f_k \hat{a}_k
\end{align}
The equation for $\hat{a}_k$ can be formally integrated and substituted in the equation for $\hat{\sigma}^- $ to find
\begin{align}
\frac{\partial }{\partial  t}  \hat{\sigma}^- = -i \omega_q \hat{\sigma}^- - i  \sum_k f_k e^{-i\omega_k (t-t_0)} \hat{a}_k(t_0) \\ \nonumber
-   \sum_k f_k^2 \int_{t_0}^t e^{-i(\omega_k) (t-\tau)} \hat{\sigma}^-(\tau) \mathrm{d} \tau.
\end{align}
We now use the Markov approximation to write $\hat{\sigma}^-(\tau) \approx  \hat{\sigma}^-(t) e^{-i(\omega_q) (\tau-t)}$, and thus
\begin{align}
\frac{\partial }{\partial  t}  \hat{\sigma}^- = -i \omega_q \hat{\sigma}^- - i  \sum_k f_k e^{-i\omega_k (t-t_0)} \hat{a}_k(t_0) \\ \nonumber
-   \sum_k  f_k^2 \left(\int_{t_0}^t e^{-i(\omega_k - \omega_q) (t-\tau)}  \mathrm{d} \tau\right) \hat{\sigma}^-(t).
\end{align}
Considering the generic equation of motion for a linearly decaying qubit, $({\partial }/{\partial  t} ) \hat{\sigma}^- = -i \omega_q \hat{\sigma}^-  - (\gamma/2)  \hat{\sigma}^-  $, we can identify real part of the last term in the equation above as the decay rate due to radiation of the qubit into the waveguide. We can extend the integral's bound to approximately evaluate this term as

\begin{align}
\gamma &= 2 \Re \left[\sum_k f_k^2  \int_{t_0}^t e^{-i(\omega_k - \omega_q) (t-\tau)}  \mathrm{d} \tau  \nonumber\right] \\ &\approx   2 \Re \left[\sum_k  f_k^2  \int_{t_0}^{\infty} e^{-i(\omega_k - \omega_q) (t-\tau)} \mathrm{d} \tau \right] \nonumber \\ &= 2\pi  \sum_k  f_k^2 \delta(\omega_k - \omega_q).
\end{align}
Assuming the coupling rate $f_k$ is a smooth function of the $k$-vector, we can evaluate this some in the continuum limit as 
\begin{align}
\gamma &= 2\pi  \sum_k  f_k^2 \delta(\omega_k - \omega_q) \\&\approx {Md} \int \mathrm{d} k  f_k^2 \delta(\omega_k - \omega_q)\\& ={L} \int \mathrm{d} \omega \left(\frac{\partial k}{\partial \omega}\right)  f_k^2 \delta(\omega_k - \omega_q) \\&= \frac{L}{c}  f(\omega_q)^2 n_g(\omega_q)  .
\end{align}
It is evident that reducing the group velocity increases the radiation decay rate of the qubit.A similar analysis can be applied to find the decay rate of a linear cavity with resonance frequency of $\omega_0$ (i.e. a harmonic oscillator) that has been coupled to the waveguide with coupling constant $g(\omega)$. In this case we find
\begin{align}
&\gamma = \frac{L}{c}  g(\omega_0)^2 n_g(\omega_0),  \nonumber\\
&Q_e = \omega_0/\gamma = \frac{\omega_0 c}{L} \frac{1}{g(\omega_0)^2 n_g(\omega_0) }.
\end{align}

\section{Disorder and Anderson localization}
\label{App:C}

Propagation of electron waves in a one dimensional quasi-periodic potential is described by
\begin{align}\label{Eq:KP}
\left[-\frac{\partial^2}{{\partial x}^2} + \sum_n (U+ U_n) \delta(x-an)\right]\psi_q(x) = q^2\psi_q(x).
\end{align}
Here, $q$ is the quasi momentum and $U_n$ is the random variable that models compositional disorder at position $x = na$. Disorder leads to localization of waves with a characteristic length defined as
\begin{align}
\ell^{-1} = \lim_{N \rightarrow \infty} \left< \frac{1}{N}  \sum_{n=0}^{N-1} \ln{\left|\frac{\psi_{n+1}}{\psi_n}\right|} \right>.
\end{align}
Here, the brackets represent averaging over different realization of the disorder, whereas the summation accounts for spatial/temporal averaging for traveling waves. For this model, previous authors have found the localization length to be \cite{HernandezHerrejon:2010ef,DossettiRomero:2004ck,Izrailev:1995by}
\begin{align}
\frac{\ell}{d} = \frac{2\Gamma(1/6)}{6^{1/3} \sqrt{\pi}} \sigma ^{-2/3}\approx 3.45\sigma ^{-2/3}.
\end{align}
In this model $\sigma^2 =  \langle U_n^2\rangle \sin^2{(q_0 a)}/q_0^2$ is a parameter that quantifies the strength of disorder, and $q_0$ is the value of quasi-momentum at the band-edge.

\begin{figure*}[t!]
\centerline{\includegraphics[width = \textwidth]{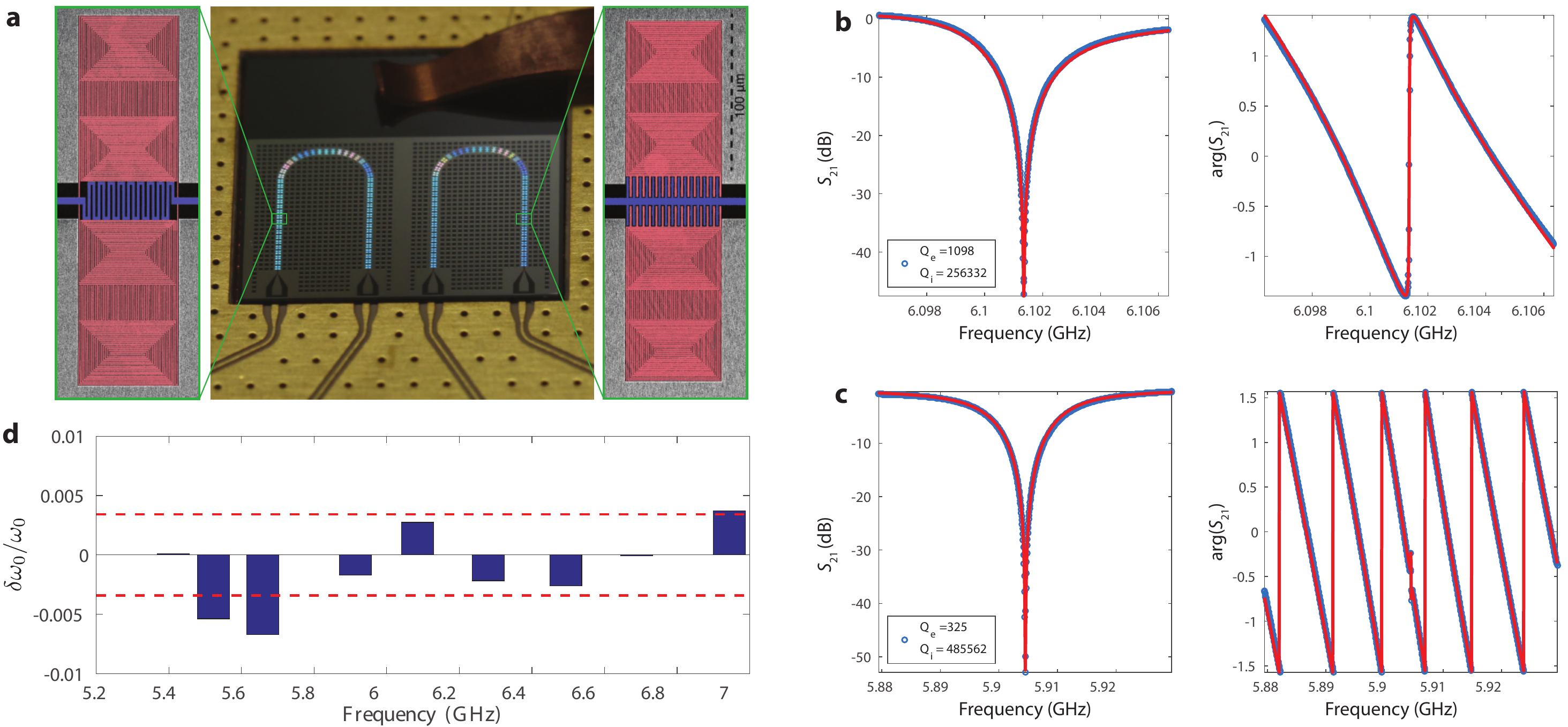}}
\caption{\textbf{a}, Optical and SEM images of microwave resonator array chip.  Middle: optical image of the chip with two arrays of coupled resonators on a $1\times1$ cm silicon chip.  Left and Right: SEM image (false-color) of the fabricated inductively (left) and capacitively (right) coupled microwave resonator pairs. The resonator region is colored red and the waveguide central conductor is colored blue. \textbf{b-c}, Amplitude and phase response of two capacitively-coupled microwave resonator pairs measured at the fridge temperature $T_f \approx 7$ mK. The legends show the intrinsic ($Q_i = \omega_0/\gamma_i$) and extrinsic ($Q_e = \omega_0/\gamma_e$) quality factors extracted from a Fano line shape fit. \textbf{c}, Statistical variations in the resonance frequency of 9 resonators with a wire width of $500$ nm. The dashed lines mark the standard deviation of the normalized error equal to $\sigma = 0.3\%$.} 
\label{fig:ResImage}
\end{figure*}

Now, we consider the propagation of current waves in a one dimensional waveguide that has been periodically loaded with resonators (a similar analysis can be applied to the voltage waves for the case of inductively coupled resonators). Starting from Eq.\,(\ref{Eq:ResDisp}), it is straightforward to find
\begin{align}
\frac{\partial^2 I(x)}{{\partial x}^2}  + I(x){\left(\frac{\omega}{c}\right)}^2 n^2 \left[1  + \sum_n\frac{d \Delta  \delta(x-an)}{\omega_{0,n} - \omega + i\gamma_i} \right]    = 0 .
\end{align}
By comparing this equation with the Schrodinger equation for the Kronig-Penny model Eq.\,(\ref{Eq:KP}) we find
\begin{align}
q^2  &\rightarrow  {\left(\frac{\omega}{c}\right)}^2 n^2\nonumber \\
U + U_n  &\rightarrow  -{\left(\frac{\omega}{c}\right)}^2 n^2 \left[\frac{d \Delta}{\omega_{0,n} - \omega + i\gamma_i}\right].
\end{align}
For small variation in resonance frequencies, $\delta \omega_{0}$, we can expand the resonance potential term to find
\begin{align}
U_n =  -{\left(\frac{\omega_0}{c}\right)}^2 n^2 \frac{\partial}{\partial \omega_{0,n}} \left(\frac{d \Delta }{\omega_{0,n} - \omega + i\gamma_i}\right) \delta \omega_{0} 
\end{align}
By evaluating the expression for $U_n$ and substituting it in the relation above for $\sigma^2$, we find

\begin{align}
\sigma^2_{\text{low}}  ={\left(\frac{\gamma_e}{\gamma_i}\right)}^4{\left(\frac{\delta \omega_0}{\Delta}\right)}^2,  \nonumber \\
\sigma^2_{\text{high}}  ={\left(\frac{\gamma_e}{\Delta}\right)}^4{\left(\frac{\delta \omega_0}{\Delta}\right)}^2 .
\end{align}

The analysis above gives us the localization length from disorder, ${\ell_\mathrm{diss}}$. In addition to disorder, the loss in the waveguide leads to an exponential extinction of the wave's amplitude. The localization length from loss, ${\ell_\mathrm{loss}}$, can be found by solving for the complex band structure and setting $\ell_\text{loss} = 1/\text{Im}(k)$. Finally, the total localization length can be found by adding the effect of both contributions as
\begin{align}
\frac{1}{\ell_\mathrm{total}} = \frac{1}{\ell_\mathrm{diss}} + \frac{1}{\ell_\mathrm{loss}}.
\end{align}

\section{Characterization of lumped-element microwave resonators}
\label{App:D}

We have achieved a characteristic size of $\lambda_0/150$ ($130 \mu$m by $76 \mu$m  for $\omega_0/2\pi \approx$ 6 GHz) and $\lambda_0/76$ ($155 \mu$m by $92 \mu$m  for $\omega_0/2\pi \approx$ 10 GHz), using a wire width of 500 nm and 1 $\mu$m, respectively. 

 Figure\,\ref{fig:ResImage} shows the typical amplitude and phase of measured for a waveguide coupled to a pair of identical resonators. Microwave spectroscopy of the fabricated resonators is performed in a dilution refrigerator cooled-down to a temperature of $T_f \approx 7$~mK. The input microwave is launched onto the chip via a 50-$\Omega$ CPW. The output microwave signal is subsequently amplified and analyzed using a network analyzer (for more details regarding the measurement setup, refer to Ref.~\cite{Dieterle:2016kj}). We have extracted the intrinsic and extrinsic decay rates of the cavity by fitting the transmission data to a Fano line shape of the form
\begin{align}
S_{21} (\omega)= 1- \frac{\gamma_e e^{i\phi_0}}{\gamma_i+\gamma_e + 2i (\omega - \omega_0)}.
\end{align} 
Here $\gamma_e $ and $\gamma_i$ are the extrinsic and intrinsic decay rates of the resonator, respectively. The phase $\phi_0$ is a parameter that sets the asymmetry of the Fano line shape \cite{Gao:2008td}.
The data demonstrates that it is possible to adjust the external coupling to the resonator in a wide range without much degradation in the internal quality factor (it is straightforward to convert the extrinsic quality factor $Q_e$ to the coupling constants $g_k$ used in our theoretical analysis above). We have compared the measured resonance frequency with the resonance frequency found from numerical simulations in Fig.\,\ref{fig:ResImage}d. We find that the measured resonance frequencies are in agreement with the simulated values, with a multiplicative scaling factor of 0.85. Using this scale factor, we have measured a random variation $0.3\%$ in the resonance frequency. It has been previously suggested that the shift in the resonance frequency and its statistical variation can be attributed to the kinetic inductance of the free charge carriers in the superconductor, and the variations can be mitigated by increasing the wire width \cite{Underwood:2012hx}.

\section{Qubit frequency shift and the Purcell-limited lifetime}
\label{App:E}

\begin{figure}[t]
\begin{center}
\ctikzset{bipoles/capacitor/height/.initial=.40}   
\ctikzset{bipoles/capacitor/width/.initial=.20}
\ctikzset{bipoles/generic/width/.initial=.25}
\ctikzset{bipoles/generic/height/.initial=.25}
\begin{circuitikz}[scale=1] 
\draw (1,0) [capacitor,*-, l_=$C_{g}$]
    to  (4,0) node[anchor=south] {};

\draw (2, -3) {}
    to [capacitor, l=$C_q$] (2, -.75) {};
    
    \draw (0, -.75) {} to (2, -.75) {}; 
     \draw (1, -.75) {} to (1, 0) node[anchor=east] {$V_{q}$}; 
        
    \draw (0, -3) {}
    to [barrier, l^=$E_{J}$] (0, -.75) {};

\draw (0, -3) {}
    to [generic] (0, -.75) {};
    
\draw (4,0) to[TL,*-*, l=metamaterial waveguide] (7,0);
\draw (4,-3) to[TL,*-*] (7,-3);    
    
\draw (3.5,1) [dashed] rectangle  (8,-4);
 
\draw [-,thick] (3.75,-0.5) to (3,-3.75);
\draw [->,thick] (3.75,-0.5) to (4.25,-0.5);

\node at (3,-4) {$ Z_\text{line}$};

\draw (7, -3) {}
to  [resistor, l_=$R_L$] (7, 0) {};

\draw (0, -3) node[anchor=east] {}
    to (4, -3) node[anchor=west] {};

\end{circuitikz}

\caption{{Circuit diagram for qubit that is capacitively coupled to a metamaterial waveguide with a resistive termination.}}\label{fig:circuit-model}
\end{center}
\end{figure}
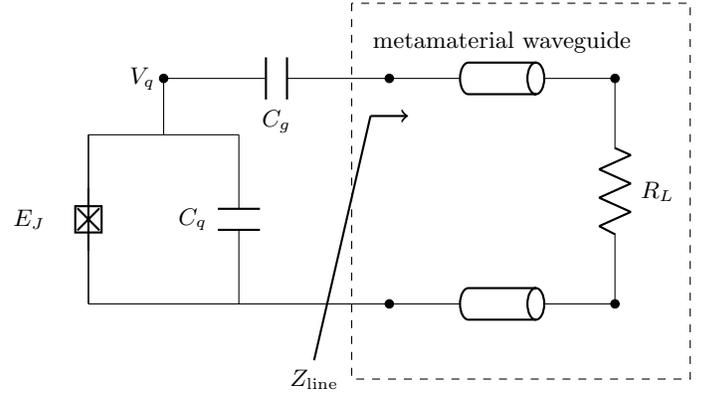

The qubit frequency shift can be derived from circuit theory by modeling the qubit as a linear resonator. Consider the circuit diagram in Fig.\,\ref{fig:circuit-model}. The load impedance seen from the qubit port can be written as
\begin{equation}
Z_L (\omega) = \frac{1}{i\omega C_g} + Z_\text{line}(\omega),
\end{equation}
and 
\begin{equation}
Y_L (\omega) = \frac{i\omega C_g}{1+ Z_\text{line}(\omega)i\omega C_g}.
\end{equation}
For weak coupling, the decay rate can be found using the real part of the load impedance as 
\begin{align}\label{Eq.kappaE}
\kappa  \simeq {\omega_q^2 L_J \Re\left[Y_{L}(\omega_q) \right]}.
\end{align}
Here, $\omega_q$ is the resonance frequency of the qubit. Similarly, the shift in qubit frequency is found as 
\begin{align}\label{Eq.FreqShift}
{\Delta \omega_q} \simeq   - \frac{ \omega_q^2 L_J }{2} \Im\left[Y_{L}(\omega_q) \right] .
\end{align}

For a transmon qubit, we have the following relation that approximate its behavior in the linearized regime 
\begin{align}
L_J = \frac{{\left( \frac{\Phi_0}{2\pi} \right)}^2}{E_J},\\
\omega_q =  \frac{1}{\sqrt{L_J C_q}}.
\end{align}
 
 We first use the simplified continuum model to find the input impedance $Z_\text{line}$
\begin{align}
Z_\text{line}(\omega)  = Z_B(\omega) \frac{ R_L+ Z_B(\omega) \tanh{\left[\Im{(k)} x \right]} }{ Z_B(\omega)+  R_L \tanh{\left[\Im{(k)} x \right]}}.
\end{align}
Here, $\Im{(k)}(\omega)$ is the attenuation constant (we are assuming  $\Re{(k)}(\omega) = 0$, i.e. valid when the value of $\omega$ is within the bandgap), $Z_B(\omega)$ is the Bloch impedance of the periodic structure, and $x$ is the length of the waveguide. Assuming $\Im{(k)} x \gg 1$, this expression can be simplified as
\begin{align}\label{Eq:Bloch}
Z_\text{line}(\omega)  \approx & Z_B(\omega)+ \frac{4R_L {| Z_B(\omega)|}^2}{{R_L}^2+ {| Z_B(\omega)|}^2} e^{-2\Im{(k)} x} \nonumber\\
 \approx &Z_B(\omega)+ 4R_L e^{-2\Im{(k)} x}.
\end{align}
Note that we have assumed $R_L \ll {| Z_B(\omega)|}$ to make the last approximation. For weak coupling, the qubit coupling capacitance, $C_g$, should be chosen such that the (magnitude of ) impedance $Z_g = 1/(i \omega C_g)$ is much larger than $|Z_\text{line}| $. In this situation, we use Eq.\,(\ref{Eq.FreqShift}) and Eq.\,(\ref{Eq:Bloch}) to find 
 \begin{align}
 \frac{\Delta \omega_q}{\omega_q} & = -\frac{1}{2}(L_J \omega_q)(C_g \omega_q)-\frac{1}{2}(L_J \omega_q){(C_g \omega_q)}^2 \text{Im}[Z_B(\omega_q)] \nonumber \\
  & = -\frac{{C_g}}{2C_q} -\frac{{C_g}}{2C_q} \text{Im}[Z_B(\omega_q)]C_g \omega_q .
 \end{align}
Note that the first term in the frequency shift is merely caused by addition of the coupling capacitor to the overall qubit capacitance.

We find the qubit's radiation decay rate by substituting Eq.\,(\ref{Eq:Bloch}) in Eq.\,(\ref{Eq.kappaE})
\begin{align}
\kappa = \frac{4\omega_q^2  {C_g}^2 }{C_q} R_L e^{-2\Im{(k)}(\omega) x}.
\end{align}
Subsequently, the radiative lifetime of the qubit can be written as
\begin{align}\label{Eq:suppLifetime}
T_\text{rad}= \frac{C_q}{4\omega_q^2  {C_g}^2 R_L}  e^{2x/\ell(\omega_q)},
\end{align}
where $\ell = 1/\Im{(k)}$ is the localization length in the bandgap. We note that the analysis from circuit theory is only valid for weak qubit-waveguide coupling rates, where the Markov approximation can be applied. In the strong coupling regime, the qubit frequency and lifetime can be found by numerically finding the zeros of the circuit's admittance function $Y = Y_L +Y_q$, where $Y_q = i\omega_q C_q + 1/(i\omega_qL_J)$.

\subsection{Group delay and the qubit lifetime profile}

Equation\,(\ref{Eq:suppLifetime}) demonstrates the relation between the qubit lifetime and the localization length. Moving the qubit frequency beyond the gap, results in a drastic increase in the localization length and subsequently reduces the qubit lifetime. The normalized slope of the lifetime profile in the vicinity of the band-edge can be written as
\begin{align}
\left|\frac{1}{T_\text{rad}} \frac{\partial  T_\text{rad}}{\partial  \omega}\right| =  \left|x \frac{\partial  \Im{(k)} }{\partial  \omega}\right| = \left|x \Im{(n_g)}/c\right|.
\end{align}
We now evaluate Eq.\,(\ref{Eq:suppGroupIndx}) to find the group index at the upper and lower band-edges $\delta_\pm = 0$ 
\begin{align}
|\Re(n_g)| = |\Im(n_g)| =
  \begin{cases}
    {n\omega_{c-}} \sqrt{\frac{\Delta}{{8\gamma_i}^3}  }        & \quad \text{for } \omega= \omega_{c-}, \\
     { {n\omega_{c+}} \frac{1 }{\sqrt{8\Delta \gamma_i}} }    & \quad \text{for }  \omega= \omega_{c+}.
  \end{cases}
\end{align}
Consequently, we can write the normalized slope of the lifetime profile at the band-edge as
\begin{align}
\left.\left(\frac{1}{T_\text{rad}} \left|\frac{\partial  T_\text{rad}}{\partial  \omega}\right|\right) \right|_{\omega = \omega_{c\pm}} = &   \left|x\Im{\left[n_g(\omega_{c\pm})\right]}/c\right|  \nonumber \\ = & \left|x \Re{\left[n_g(\omega_{c\pm})\right]}/c\right| \nonumber \\ = & \tau_\text{delay}.
\end{align}
This result has a simple description:  the normalized slope of the lifetime profile at the band-edge is equal to the (maximum) group delay. 

\end{document}